\documentclass[aps,twocolumn,pra,showpacs,floatfix]{revtex4}
\usepackage{epsfig}
\usepackage{graphicx}
\usepackage{dcolumn}
\usepackage{amssymb,amsmath}

\begin{document}

\title{Chaos-induced enhancement of resonant multielectron recombination in
highly charged ions: Statistical theory}

\author{V. A. Dzuba$^1$, V. V. Flambaum$^1$, G. F. Gribakin$^2$,
and C. Harabati$^1$}

\affiliation{$^1$School of Physics, University of New South Wales, Sydney 2052,
Australia}
\affiliation{$^2$Department of Applied Mathematics and Theoretical Physics,
Queen's University, Belfast BT7 1NN, Northern Ireland, UK}

\date{\today}
\begin{abstract}
A statistical theory of resonant multielectron recombination based on
properties of chaotic eigenstates is developed. The level density of
many-body states increases exponentially with the number of excited electrons.
When the residual electron-electron interaction exceeds the interval between
these levels, the eigenstates (called compound states or compound  resonances
if these states are in the continuum)  become ``chaotic'' superpositions of
large numbers of Hartree-Fock configurational basis states. This situation
takes place in some rare-earth atoms and many open-shell multiply charged ions
excited in the process of electron recombination. Our theory describes resonant
multielectron recombination via dielectronic doorway states
leading to such compound resonances. The result is a radiative capture
cross section averaged over a small energy interval containing several compound
resonances. In many cases individual resonances are not resolved
experimentally (since the interval between them is small, e.g., $\le 1$ meV,
possibly even smaller than their radiative widths),
therefore, our statistical theory should correctly describe the experimental
data. We perform numerical calculations of the recombination cross sections for
tungsten ions W$^{q+}$, $q=18$--25. The recombination rate for W$^{20+}$
measured recently [Phys. Rev. A {\bf 83}, 012711 (2011)] is $10^3$ greater
than the direct radiative recombination rate at low energies, and our result
for W$^{20+}$ agrees with the measurements.
\end{abstract}

\maketitle

\section{Introduction}

The majority of atoms in highly excited states and some open-shell (e.g.,
rare-earth) atoms even in vicinity of the ground state behave as complex,
chaotic many-electron systems with very dense spectra and strong configuration
mixing. The many-electron wave function in such a system is a mixture of a
large number of many-excited-electron basis states (Slater determinants)
with nearly random mixing coefficients. Ordinary theoretical methods, such as
configuration interaction, which work well for atomic systems with few (say,
two) electrons above closed subshells, become practically useless here. On the
other hand, such complex systems can be described using statistical approaches.
Statistical methods for chaotic compound states are widely used in nuclear
physics (see, e.g., Refs.~\cite{Bohr:69,Flambaum:93,FG95,Zel,FG2000}). Similar
methods were developed for open-shell atomic systems
\cite{Ce,Gribakin99,Flambaum02,Gribakin03}. In Ref.\cite{Ce} we used the
cerium atom as a testing ground for the applications of the statistical theory,
and studied properties of the Hamiltonian matrix in chaotic many-body systems,
statistics of the intervals between energy levels, average orbital occupation
numbers as functions of the excitation energy, electromagnetic amplitudes
between chaotic many-body states, enhancement of weak interaction effects
in such states, and electronic and electromagnetic widths of chaotic compound
resonances. The statistical theory was later also tested for multicharged
ions~\cite{Gribakin99,Flambaum02,Gribakin03}.

In the present paper we want to use the statistical theory to describe the
effect of compound resonances on electron recombination with open-shell
multicharged ions. Electron recombination is an important process in
laboratory and cosmic plasmas, as well as in ion storage rings. Theory and
experiment agree very well for relatively simple systems with one or two
valence electrons above closed shells (see, e.g., Ref.~\cite{Tokman02}).
For more complex systems theory and experiment often deviate
significantly. For example, strong enhancement of the recombination
rate were observed for Au$^{25+}$~\cite{Hoffknecht:98}, U$^{28+}$~\cite{U28} and
W$^{20+}$~\cite{schippers11} at low electron energies. For such ions the
observed recombination rate is orders of magnitude greater than that
due to direct radiative recombination (RR). In simpler systems
the enhancement is due to resonant dielectronic recombination, though even for
ions like Fe$^{9+}\,3p^5$ and Fe$^{9+}\,3p^4$ the dielectronic
recombination appears to be deficient \cite{Lestinsky09}.
In complex open-shell ions such as Au$^{25+}\,4f^8$ and isoelectronic W$^{20+}$
the absolute majority of the resonances correspond to many-excited-electron
eigenstates which have very high density (exponentially small level spacings).
In Refs. \cite{Gribakin99,Flambaum02} we used a statistical
approach to show that the 200-times enhancement over RR observed in Au$^{25+}$
\cite{Hoffknecht:98} is due to electron capture in these compound
resonances.

In this paper we use the statistical theory to calculate the recombination rates
for tungsten ions W$^{q+}$, $q=18$--25. Our results for W$^{20+}$, where the
measured rate at low ($\sim 1$~eV) energies is $10^3$ times  higher
than the radiative rate \cite{schippers11}, are in good agreement with
experiment. For other tungsten ions we predict similar strong enhancements of
the recombination rate.

\begin{figure*}[t!]
\centering
\includegraphics*[width=0.8\textwidth]{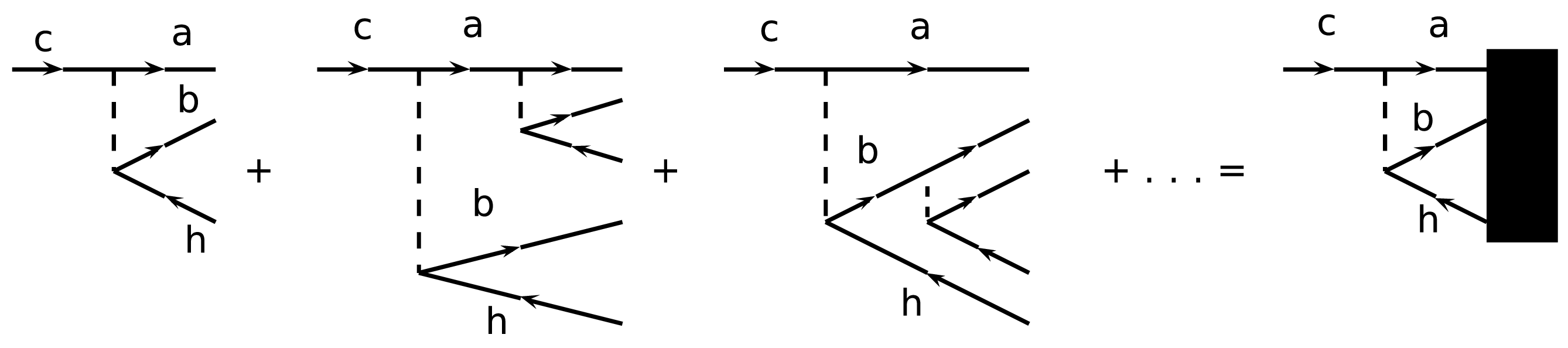}
\caption{Schematic diagram of electron capture into a strongly mixed
multiconfigurational eigenstate (dark block) through a dielectronic doorway
configuration $h^{-1}ab$. Dashed lines show the Coulomb interactions.}
\label{fig:diagram}
\end{figure*}

A detailed derivation of the statistical theory for the recombination cross
section is presented in the next section.
Numerical calculations based on our statistical theory are somewhat similar to
those of standard dielectronic recombination. However, in the statistical
theory one does not need to diagonalize potentially very large Hamiltonian
matrices for the excited states of the compound ion. Instead, the statistical
formalism contains new parameters, namely the spreading width $\Gamma_{\rm spr}$
and average electron orbital occupation numbers in the initial (ground) state
of the ion. The total cross section of electron capture into compound
resonances is given by
\begin{eqnarray}\label{sigma}
\bar \sigma _c&=&\frac{\pi ^2}{2k^2}\sum _{a b h ,lj}
\Gamma_{\rm spr}\frac{ | \langle a,b |\hat v |h, \varepsilon lj\rangle  
-\langle b, a |\hat v |h,\varepsilon lj\rangle |^2}
{(\varepsilon -\varepsilon _a -
\varepsilon _b +\varepsilon _h )^2 +\Gamma_{\rm spr}^2/4}  \nonumber \\
&\times &\langle \hat n_h \hat n_c (1-\hat n_a)(1-\hat n_b)\rangle _i
\end{eqnarray}
where $\varepsilon $ and $k$ are the energy and wavenumber of the incident
electron, $ \langle a,b |\hat v |h, \varepsilon lj\rangle $ and
$ \langle b, a |\hat v |h, \varepsilon lj \rangle $ are the direct and
exchange Coulomb matrix elements, respectively, the sum is over the
single-particle states (orbitals) of the hole ($h$) and excited electrons ($a$
and $b$), and the partial waves $lj$ of the incident electron ($c$), and
$\hat n_a $, $\hat n_b $, etc., are the corresponding occupation numbers. The
expectation value
$ \langle \hat n_h \hat n_c (1-\hat n_a)(1-\hat n_b)\rangle _i $ over the
initial target state tells one that to transfer electrons from orbitals $h$ and
$c$ into $a$ and $b$, the former must be at least partially occupied and the
latter at least partially empty. The continuum electron wave function $c$
is normalized to the delta-function of energy and $\hat n_c=1$.

The presence of the
spreading width $\Gamma_{\rm spr}$ in Eq.~(\ref{sigma}) is due to the fact that
the two-electron-one-hole excitation $h^{-1}ab$ is not an eigenstate of the
highly excited ion. This state is embedded in the dense spectrum of other,
multiply-excited states and is strongly mixed with them. This dielectronic
state plays the role of a {\em doorway} state which ``decays'' to more
complicated states, and the  width of this ``internal'' decay is denoted
by $\Gamma_{\rm spr}$. This process is faster than either autoionization or
radiative decay (``external'' decays), and $\Gamma_{\rm spr}$ is greater than
the  autoionization or radiative widths (see below).

Figure~\ref{fig:diagram} presents a perturbative, diagrammatic picture of
electron capture through the doorway $h^{-1}ab$. In this temporal picture
the process looks as a series of electron collisions. The initial electron
$c$ collides with an atomic electron in state $h$ and excites it into state $b$.
Then one of the excited electrons interacts with another atomic electron to
produce more excitations, etc. As a result, the initial electron energy is
shared between many electrons, and none of them has enough energy to escape.
In this way a long-lived compound resonance is formed. A similar 
picture of neutron capture by nuclei dates back to Niels Bohr \cite{Bohr:36}
and is very well known to nuclear physicists. This temporal picture assumes
that individual time steps $\delta t$ can be resolved. However, the uncertainty
relation, $\delta t \delta E \gtrsim \hbar $, would then require a large
energy uncertainty $\delta E$. In the recombination process the total energy
of the system ``electron $+$ ion'' is well defined. This means that all the
components (steps in the process in Fig.~\ref{fig:diagram}) are present in
the long-lived quasistationary compound state which captures the electron.
The language of strong  configuration mixing is more appropriate in this case.
With a perfect energy resolution one would see a very dense spectrum of narrow,
possibly overlapping resonances. Broad doorway dielectronic states (with
width $\Gamma_{\rm spr}$) can only introduce a variation of the average height
of these narrow compound resonances on the energy scale
$\Delta \varepsilon \sim \Gamma_{\rm spr}$.

Compared with the total resonant capture cross section, the recombination
cross section
\begin{equation}\label{sigmarad}
\bar \sigma _r=\omega_f  \bar \sigma _c
\end{equation}
contains an additional factor $\omega_f $, known as the fluorescence yield.
It accounts for the probability of radiative stabilization of the resonances
(as opposed to autoionization), and is given by
\begin{equation}\label{rad}
\omega_f=\frac{\Gamma^{(r)}}{\Gamma^{(r)}+\Gamma^{(a)}},
\end{equation}
where $\Gamma^{(r)}$ and $\Gamma^{(a)}$ are the resonance radiative and
autoionization widths, respectively. Expressions for these widths as well as
the spreading width  $\Gamma_{\rm spr}$, are presented in the next section.

Note that the  capture cross section (\ref{sigma}) is not very sensitive to
the specific value of the spreading width $\Gamma_{\rm spr}$ which for
multicharged ions is about $0.5$ a.u. \cite{Gribakin99}
(see Table \ref{t:W}). After angular reduction of the Coulomb matrix elements
in Eq. (\ref{sigma}) [Sec. \ref{theory}, Eq.~(\ref{eq:capture-explicit})],
numerical calculations of the capture cross section are straightforward.  

An additional simplification occurs in heavy open-shell ions like
Au$^{25+}$ and W$^{20+}$, which have almost unit fluorescence yield.
Indeed, the compound states in these ions contain very large numbers of
principal basis-state components, $N \sim 10^4$. Each component contributes
to the radiative decay into a large number of states below this compound state.
In contrast, only one or few dielectronic (doorway) state components have
nonzero Coulomb matrix elements that allow electron autoionization
into the continuum (see Fig.~\ref{fig:diagram} and Eq.~(\ref{sigma})).
Therefore,  the autoionization width of the compound resonance is suppressed
by the small weight factor $1/N$ of  the dielectronic components in the
compound state. This means  that the captured low-energy electron cannot
escape, i.e., after the capture the radiative process happens with nearly
100\% probability. (A similar effect in neutron capture by nuclei is described,
e.g., in Ref.~\cite{FS84}).  
In this situation $\Gamma^{(a)}\ll \Gamma^{(r)}$ and $\omega_f \approx 1$, so
that the electron recombination cross section, Eq. (\ref{sigmarad}),
is independent of the radiative width. In this regime one observes maximum
chaos-induced enhancement of the resonant multielectron recombination.
For example, the electron capture cross section calculated using
Eq.~(\ref{sigma}) for Au$^{25+}$ \cite{Flambaum02}, was found to be in good
agreement with experiment at low energies.
On the other hand, in ions with a smaller number of active electrons,
the number of components $N$ may not be so large, leading to
$\Gamma^{(a)}>\Gamma^{(r)}$ and $\omega_f < 1$.

\section{Theory}\label{theory}

\subsection{Resonant recombination cross section}

The resonant radiative electron-ion recombination cross section is given by
the sum over the resonances $\nu$ with the angular momentum and parity $J^\pi$,
\begin{equation}\label{eq:sigmar}
\sigma _r=\sum_{\nu}g(J)\sigma_\nu,
\end{equation}
where $g(J)=(2J+1)/[2(2J_i+1)]$ is the probability factor due to random
orientation of the electron spin and angular momentum $J_i$ of the target
ion \cite{Landau}. The individual resonant contributions $\sigma_\nu$ are
given by the Breit-Wigner formula \cite{Landau},
\begin{equation}\label{cross-section}
\sigma_{\nu}  = \frac{\pi}{k^2} \frac{\Gamma_\nu^{(a)}\Gamma_\nu^{(r)}}
{(\varepsilon-\varepsilon_\nu)^2+\Gamma_\nu^2/4},
\end{equation}
where $\Gamma_\nu^{(r)}$ is the the radiative decay rate (the total ``inelastic
width''), $\Gamma_\nu^{(a)}$ is the autoionization decay rate (the ``elastic
 width''), and $\Gamma _\nu =\Gamma_\nu^{(r)}+\Gamma_\nu^{(a)}$ is the total
width of the level $\nu$. (We assume here that other inelastic channels, e.g.,
electronic excitation, are closed, which is correct at low incident electron
energies). The energy of the $\nu$th resonance is given relative
to the ionization threshold $I$ of the final compound ion,
$\varepsilon_\nu=E_\nu-I $. 

For systems with dense compound resonance spectra the recombination cross
section displays rapid energy dependence, that may not even be resolved
experimentally. Thus it is natural to average the cross section over an
energy interval $\Delta \varepsilon $ which is large compared with the small
mean level spacing $D_{J^\pi}$ and the total resonance width, but much smaller
than $\varepsilon $. This gives
\begin{eqnarray}\label{aDRcs}
\frac{1}{\Delta \varepsilon} \int \sigma_\nu d\varepsilon=
\frac{2\pi^2}{k^2}\frac{ \Gamma^{(r)}_\nu  \Gamma ^{(a)}_\nu }
{\Delta \varepsilon \Gamma _\nu },
\end{eqnarray}
where the integration limits are formally infinite, since the contribution of
each resonance to cross section decreases rapidly away from
$\varepsilon \approx \varepsilon_\nu$. The number of resonances with a given
$J^\pi$ within
$\Delta \varepsilon $ is $\Delta \varepsilon/D_{J^\pi} $, and after averaging, 
the recombination cross section (\ref{eq:sigmar}) becomes
\begin{equation}\label{aDRcs-final}
\bar \sigma _r=\frac{\pi^2}{k^2}\sum_{J^\pi}\frac{2J+1}{(2J_i+1)D_{J^\pi}}
\left\langle \frac{ \Gamma^{(r)} \Gamma ^{(a)}}{\Gamma}\right\rangle .
\end{equation} 
Here $\langle \dots \rangle $ means averaging of the width factor at the given
energy.

If the fluorescence yield $\omega_f=\Gamma ^{(r)}/\Gamma$ fluctuates weakly
from resonance to resonance, the recombination cross section
$\bar \sigma _{r}$ can be factorized, i.e.,
$\bar \sigma _{c}^r=\omega_f\bar\sigma _{c}$. For
$\omega_f\approx 1$ the energy-averaged capture cross section
\begin{equation}\label{aCapcs}
\bar \sigma _{c}=\frac{\pi^2}{k^2}\sum_{ J^\pi}\frac{(2J+1)}{(2J_i+1)}
\frac{\langle \Gamma ^{(a)}\rangle } {D_{J^\pi}}
\end{equation} 
is the same as the recombination cross section.

In the opposite case of small radiative widths, autoionization dominates
($\Gamma^{(r)}\ll \Gamma^{(a)}$) and $\omega_f\ll 1$, so Eq. (\ref{aDRcs-final})
yields the recombination cross section in the form
\begin{equation}\label{radiative-final}
\bar \sigma _r^a=\frac{\pi^2}{k^2} \frac{\langle \Gamma^{(r)}\rangle }
{(2J_i+1)}\sum_{J^\pi}(2J+1)\rho_{J^\pi},
\end{equation}
where $ \rho_{J^\pi}=1/D_{J^\pi}$ is the level density and
$\langle \Gamma^{(r)}\rangle $ is given by Eq. (\ref{eq:Gamma_r}). Note that
the sum in Eq.~(\ref{radiative-final}) is the total density of states, which
can be found without constructing states with definite $J$.

To estimate the recombination cross section in the general case one can use
the following formula,
\begin{equation}\label{eq:rec_app}
\bar \sigma _r\approx \frac{\bar \sigma _c\bar \sigma _r^a}
{\bar \sigma _c+\bar \sigma _r^a}.
\end{equation}
It follows from Eqs.~(\ref{aDRcs-final})--(\ref{radiative-final}) if $\omega _f$
does not depend on $J$.

\subsection{Nature of chaotic compound states}\label{subsec:chaos}

The density of excited states $\rho(E)$ in a many-electron ion, especially with
an open shell, increases rapidly (exponentially) as the number of excited
electrons increase. Consider $n$ electrons that can be distributed among a
number of single-electron states $g=\sum _l2(2l+1)$, where $l$ is the orbital
angular momentum of the subshells available. The total number of many-body
states that can be constructed is given by
\begin{equation}\label{num-manybody}
\frac{g!}{n!(g-n)!}\approx \frac{ \exp[ n \ln(g/n)+n]}
{\sqrt{2\pi n}},
\end{equation}
where we used the Stirling formula and assumed $g\gg n$.

\begin{figure}[t!]
\centering
\includegraphics*[width=0.48\textwidth]{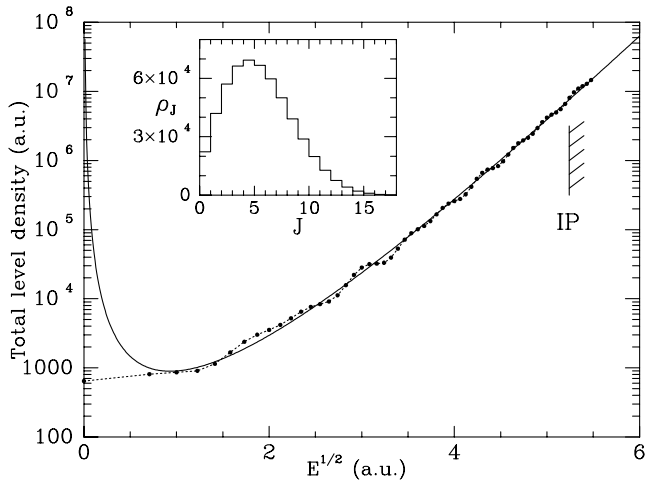}
\caption{Level density in Au$^{24+}$. The black dots are the result of the
numerical calculation \cite{Gribakin99}. The solid line is the analytical fit,
$\rho (E)=AE^{-\nu }\exp (a \sqrt{E})$, motivated by the level density
calculated using the Fermi gas model \cite{Bohr:69,Gribakin99}. The inset shows
the densities of states with different $J$ at the ionization energy $E=I$.}
\label{fig:density}
\end{figure}

Equation (\ref{num-manybody}) indicates exponential increase of the number of
many-electron states and the corresponding decrease of the energy interval
between them as the number of ``active'' electrons $n$ increases. For
example, Fig.~\ref{fig:density} shows how the density of
multielectron excited states of Au$^{24+}$ increase with energy $E$. The small
level spacings between the states mean that even a small residual
electron-electron interaction will cause strong non-perturbative mixing of the
many-electron configuration basis states (Slater determinants)
$|\Phi_k\rangle $. This occurs when the off-diagonal matrix elements of the
Hamiltonian $H_{ij}$ become greater than the energy spacing $D_{ij}$ between
the basis states $i$ and $j$ coupled by the residual interaction,
$H_{ij}> D_{ij}$.

When the mixing is strong, each eigenstate 
\begin{equation}\label{eq:eigenket}
|\Psi_\nu\rangle = \sum_k C_k^{(\nu)} |\Phi_k\rangle . 
\end{equation} 
contains a large number $N$ of principal components $|\Phi_k\rangle $. These
are the basis states for which the expansion coefficients (which satisfy
the normalization condition $\sum_k |C_k^{(\nu)}|^2=1$) have typical values
$C_k^{(\nu)}\sim 1/ \sqrt{N}$. The number of principal components in an
eigenstate can be estimated as $N\sim\Gamma_{\rm spr}/D$, where
\begin{equation}\label{gammaspread}
\Gamma_{\rm spr} \simeq \frac{2\pi \overline{H_{ij}^2}}{D}
\end{equation}  
is the spreading width, and $D$ is the mean level spacing between the basis
states (or eigenstates). Such eigenstates are called compound states, and are
well known, e.g., in nuclear physics literature \cite{Bohr:69}.

\begin{figure}[t!]
\centering
\includegraphics*[width=0.48\textwidth]{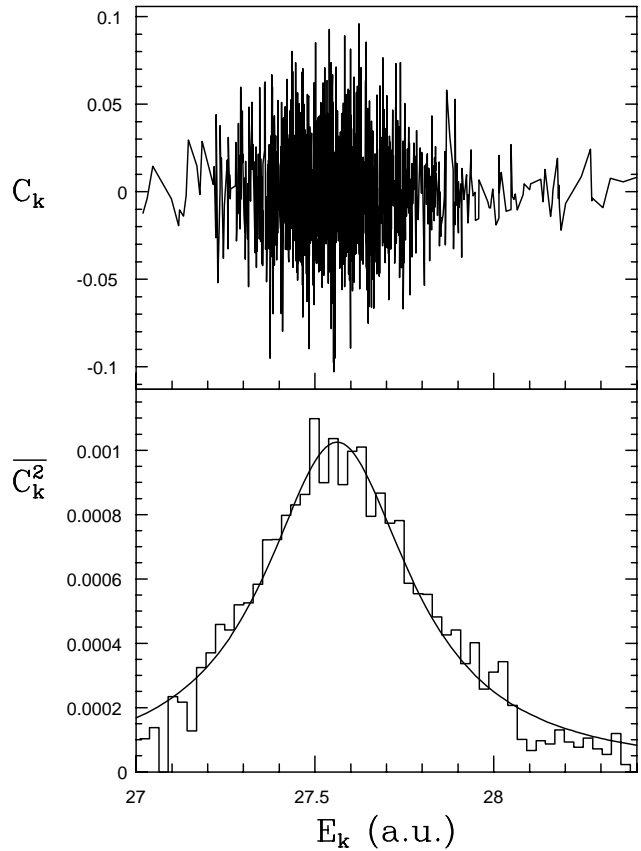}
\caption{Components of the 590th eigenstate with $J^\pi=13^-/2$ in Au$^{24+}$
from a two-configuration calculation \cite{Gribakin99} and
$\overline{C_k^2}(E)$ (histogram) fit by the Breit-Wigner formula,
Eq.~(\ref{C^2}) (solid line).
The two configurations, $4f_{5/2}^34f_{7/2}^35p_{1/2}5p_{3/2}5f_{7/2}$ and 
$4f_{5/2}^34f_{7/2}^35p_{1/2}5d_{3/2}5g_{7/2}$, produce a total of 143360
many-electron states with $J$ from $\frac{1}{2}$ to $\frac{35}{2}$.}
\label{fig:components}
\end{figure}

For example, in Au$^{24+}$ the mean spacing between the excited states with a
given angular momentum and parity, near the ionization threshold is
$D_{J^\pi}\sim 1$ meV and $\Gamma_{\rm spr} \sim 10$ eV, so that
$N\sim\Gamma_{\rm spr}/D\sim 10^4$ \cite{Gribakin99}. Numerical calculations
involving a relatively small number of configurations confirm that in this
case the eigenstates are indeed chaotic superpositions of the basis states, see
Fig.~\ref{fig:components}.  

The energies $E_k$ of the principal basis components lie within the spreading
width of the eigenenergy $E_\nu$ of the compound state,
$|E_k-E_\nu|\lesssim \Gamma_{\rm spr}$. The components outside the spreading
width decrease quickly, so that they do not give much contribution to the
normalization. It was tested in Refs. \cite{Ce,Gribakin99} that components
of the chaotic eigenstates have the statistics of Gaussian random variables
with zero mean (Fig. \ref{fig:components}, top). On the other hand, the
variation of their mean-squared value as a function of energy
(Fig. \ref{fig:components}, bottom) can be approximated by a simple
Breit-Wigner profile,
\begin{equation}\label{C^2}
\overline {\bigl| C_k^{(\nu)}\bigr|^2}=N^{-1}\frac{\Gamma_{\rm spr}^2/4}
{(E_k-E_\nu ) ^2+\Gamma_{\rm spr} ^2/4},
\end{equation}
with $N=\pi \Gamma _{\rm spr}/2D$ fixed by normalization
$$\sum _k\bigl|C_k^{(\nu)}\bigr|^2\simeq
\int \overline {\bigl| C_k^{(\nu)}\bigr|^2} dE_k/D=1.$$

To summarize, the chaotic compound states have the following properties:
(i) Each eigenstate contains a large number $N$ of principal components
$C^{(\nu )}_k\sim 1/\sqrt{N}$, corresponding to the basis states
$|\Phi _k\rangle $ which are strongly mixed together. (ii) Owing to
the strong mixing, the only good quantum numbers that can be used to classify
the eigenstates, are the exactly conserved total angular momentum and parity
$J^\pi $ and the energy. (iii) The degree of mixing in this regime is in some
sense complete, i.e., all basis states that can be mixed (within a
certain energy range) are mixed. The notion of configurations based
on the single-particle orbitals becomes largely irrelevant for the purpose of
classifying the eigenstates. Each eigenstate contains substantial
contributions of all nearby configurations.

These properties of chaotic compound states enable one to calculate the
mean-squared matrix elements of different  operators without diagonalization
of prohibitively large configuration-interaction Hamiltonian matrices.

\subsection{Mean-squared matrix elements between compound states}

Consider a two-body operator (e.g., the Coulomb interaction)
$$\hat V=\frac{1}{2}\sum_{abch} \langle ab|\hat v |hc\rangle 
a^{\dagger}_{a}a^{\dagger}_{b}a_{h}a_{c}.$$
A matrix element of $\hat V $ between two compound states, $|\Psi_\nu \rangle $
and $|\Psi_i \rangle $, is given by (see Eq. (\ref{eq:eigenket}))
\begin{equation}\label{Coulomb-matrix2}
\langle \Psi _\nu |\hat V |\Psi_i \rangle=\sum_{kk'}C_k^{(\nu)\ast}
C_{k'}^{(i)}\langle \Phi_k |\hat V |\Phi_{k'} \rangle ,
\end{equation}
or 
\begin{eqnarray}\label{Coulomb-matrix}
\langle \Psi _\nu |\hat V |\Psi_i \rangle &=& \frac{1}{2}\sum_{(ab)ch}
(\langle ab |\hat v | hc\rangle-\langle ba |\hat v |hc\rangle)\nonumber \\
&\times&\langle\Psi _\nu |a^{\dagger}_{a}a^{\dagger}_{b}a_{h}a_{c}|\Psi_i \rangle, 
\end{eqnarray}
where in the last equation the sum is restricted to distinct pairs $(ab)$,
and $\langle\Psi _\nu |a^{\dagger}_{a} a^{\dagger}_{b}a_{h}a_{c}|\Psi_i \rangle $
determines the contribution of the two-particle transition $ch\rightarrow ab$.
Due to the assumption that the expansion coefficients for chaotic compound
states are random and uncorrelated ($\overline{C_k^{(\nu)}}=\overline{{C_k^{(\nu)}}^*C_{k'}^{(i)}}=0 $ for $\nu \neq i$), the value of
the matrix element averaged over many compound states $\nu $ is zero,
\begin{equation}
\overline{\langle \Psi _\nu |\hat V |\Psi_i \rangle}=0,
\quad \overline{\langle\Psi _\nu |a^{\dagger}_{a}
a^{\dagger}_{b}a_{h}a_{c}|\Psi_i \rangle}=0.
\end{equation}

To determine the autoionization width (Sec. \ref{subsec:aw}) we need to
calculate the {\em mean-squared} matrix element. It is derived using the
statistical properties of the expansion coefficients,
$\overline{C_k^{(\nu)*}C_{l}^{(\nu)}}=\overline {| C_k^{(\nu)}|^2}\delta_{kl}$,
so that 
\begin{eqnarray}
\overline{\langle\Psi _\nu |a^{\dagger}_{a}
a^{\dagger}_{b}a_{c}a_{h}|\Psi_i \rangle \langle\Psi _i | a^{\dagger}_{c'}a^{\dagger}_{h'}a_{b'}a_{a'}|\Psi_\nu \rangle }= \nonumber \\
\delta_{aa'}\delta_{bb'}\delta_{cc'}\delta_{hh'}\overline{|\langle\Psi _\nu |a^{\dagger}_{a}a^{\dagger}_{b}a_{h}a_{c}|\Psi_i \rangle|^2}.
\end{eqnarray}
Hence, the mean-squared matrix element is
\begin{eqnarray}\label{eq:MSM}
\overline{|\langle \Psi _\nu |\hat V |\Psi_i\rangle |^2} &=&\frac{1}{4}
\sum_{abch} | \langle ab |\hat v |hc\rangle-\langle ba |\hat v |hc\rangle |^2
\nonumber \\
& \times & \overline {|\langle \Psi_\nu |a^{\dagger}_{a}a^{\dagger}_{b}
a_{h}a_{c}|\Psi_i \rangle |^2}.
\end{eqnarray}

Let us introduce the strength function
\begin{equation}\label{def:w}
w(E_k;E_\nu ,\Gamma _{\rm spr},N)\equiv \overline{C_k^{(\nu)2}},
\end{equation}
which describes the spreading of the component $k$ over the eigenstates
$\nu$ ($C_k^{(\nu)}$ are assumed to be real). This function depends on the
number of principal components $N$ of the eigenstate, Eq. (\ref{eq:eigenket}),
the spreading width $\Gamma_{\rm spr}$, and the difference $E_\nu-E_k$ between
the energies of the compound state and component $k$. In the simplest model
\cite{Bohr:69} $w(E_k;E_\nu ,\Gamma_{\rm spr},N)$ is a Breit-Wigner function, cf.
Eq.~(\ref{C^2}). Using Eq. (\ref{Coulomb-matrix2}), we then obtain
\begin{eqnarray}\label{mean-weight}
\overline {|\langle \Psi_\nu |a^{\dagger}_{a}a^{\dagger}_{b}a_{h}a_{c}|\Psi_i \rangle |^2}  =\sum_{kk'}\sum_{ll'}\overline{C_k^{(\nu)}C_{l}^{(\nu)}}~
\overline{C_{k'}^{(i)}C_{l'}^{(i)}}\nonumber \\
\times \langle\Phi _k |a^{\dagger}_{a}a^{\dagger}_{b}a_{h}a_{c}|\Phi_{k'} \rangle
\langle\Phi _{l'} |a^{\dagger}_{c}a^{\dagger}_{h}a_{b}a_{a}|\Phi_l \rangle
\nonumber \\
 = \sum_{kk'}w_i(E_{k'})w_\nu(E_k)\hspace{28pt}\nonumber \\
\times \langle\Phi _{k'} |a^{\dagger}_{c}a^{\dagger}_{h}a_{b}a_{a}|\Phi_k \rangle
\langle\Phi _k |a^{\dagger}_{a}a^{\dagger}_{b}a_{h}a_{c}|\Phi_{k'} \rangle.
\end{eqnarray}
To obtain the last expression we used the properties of the components and the
definition (\ref{def:w}), and denoted
$w_i(E_{k'})\equiv w(E_{k'};E_i,\Gamma_{\rm spr}^{(i)},N_i)$ and 
$w_\nu (E_{k})\equiv w(E_{k};E_\nu,\Gamma_{\rm spr}^{(\nu)},N_\nu)$.

We can assume, without the loss of generality, that the number of principal
components $|\Phi_k \rangle$ in state $\nu $ is greater than or equal to the
number of components $|\Phi_{k'} \rangle$ of state $i$, i.e.,
$\Gamma_{\rm spr}^{(\nu)}/D_\nu\geq \Gamma_{\rm spr}^{(i)}/D_i$. The matrix element
$ \langle\Phi _k |a^{\dagger}_{a}a^{\dagger}_{b}a_{h}a_{c}|\Phi_{k'} \rangle $
does not vanish only if
$ |\Phi_{k}\rangle=a^{\dagger}_{a}a^{\dagger}_{b}a_{h}a_{c} |\Phi_{k'} \rangle $,
so that $ E_k-E_{k'}\simeq \varepsilon_a+ \varepsilon_b- \varepsilon_h- \varepsilon \equiv \omega_{ab,ch}$. Using closure to sum over $k$ in
Eq.~(\ref{mean-weight}), we obtain
\begin{eqnarray}\label{mean-weight2}
\overline {|\langle \Psi_\nu |a^{\dagger}_{a}a^{\dagger}_{b}a_{h}a_{c}|\Psi_i \rangle |^2}
 = \sum_{k'}w_i(E_{k'})w_\nu(E_{k'}+\omega_{ab,ch})\nonumber \\
\times \langle\Phi _{k'} |\hat{n}_h\hat{n}_c(1-\hat{n}_a)(1-\hat{n}_b)|\Phi_{k'} \rangle.
\end{eqnarray}
In deriving this equation we used the anticommutation relations
satisfied by the creation and annihilation operators, and introduced the the
occupation number operators $ \hat{n}_a=a^{\dagger}_aa_a $. The matrix element
$ \langle\Phi _{k'} |\hat{n}_h\hat{n}_c(1-\hat{n}_a)(1-\hat{n}_b)|\Phi_{k'} \rangle $
is equal to unity if the orbitals $h$ and $c$ are occupied, while the orbitals
$a$ and $b$ are vacant in the state $|\Phi_{k'} \rangle $, i.e., the
transition $ch\rightarrow ab$ is possible.

If one assumes that the single-electron-state occupancies vary slowly
with the excitation energy, then the matrix element of the operator
$\hat{n}_h\hat{n}_c(1-\hat{n}_a)(1-\hat{n}_b)$ in Eq.~(\ref{mean-weight2}) can
be replaced by its expectation value,
\begin{eqnarray}\label{eq:emptiness}
\sum_{k'} w_i(E_{k'})\langle\Phi _{k'} |\hat{n}_h\hat{n}_c(1-\hat{n}_a)(1-\hat{n}_b)|\Phi_{k'} \rangle =\nonumber \\
\langle\hat{n}_h\hat{n}_c(1-\hat{n}_a)(1-\hat{n}_b)\rangle_i ,
\end{eqnarray} 
subject to the normalization condition $ \sum_{k'} w_i(E_{k'})=1 $. The
right-hand side of Eq. (\ref{eq:emptiness}) is the value of the occupancy
times ``emptiness'' in the compound state $|\Psi_i \rangle$, averaged over
a number of neighboring states.

Replacing the matrix element $\langle \Phi _{k'}|\dots |\Phi _{k'}\rangle $ by
its average (\ref{eq:emptiness}) in Eq. (\ref{mean-weight2}), and  changing
summation to integration, one obtains
\begin{eqnarray}\label{mean-weight3}
\overline {|\langle \Psi_\nu |a^{\dagger}_{a}a^{\dagger}_{b}a_{c}a_{h}|
\Psi_i \rangle |^2}
= \langle\hat{n}_h\hat{n}_c(1-\hat{n}_a)(1-\hat{n}_b) \rangle_i \nonumber \\
\times\int w_i(E_{k'})w_\nu(E_{k'}+\omega_{ab,ch})\frac{dE_{k'}}{D_i}.
\end{eqnarray}
This result can be written in the following form:
\begin{eqnarray}\label{mean-weight4}
\overline {|\langle \Psi_\nu |a^{\dagger}_{a}a^{\dagger}_{b}a_{h}a_{c}|
\Psi_i \rangle |^2} = \langle\hat{n}_h\hat{n}_c(1-\hat{n}_a)(1-\hat{n}_b) 
\rangle_i \nonumber \\
\times D_\nu\tilde{\delta}(\Gamma_{\rm spr}^{(i)},\Gamma_{\rm spr}^{(\nu)},\Delta).
\end{eqnarray}
In this expression
\begin{eqnarray}\label{def:delta}
\tilde{\delta}(\Gamma_{\rm spr}^{(i)},\Gamma_{\rm spr}^{(\nu)},\Delta)
\equiv \frac{1}{D_\nu}\int w(E_{k'};E_i,\Gamma_{\rm spr}^{(i)},N_i)\nonumber \\
\times w(E_{k'}+\omega_{ab,ch};E_\nu,\Gamma_{\rm spr}^{(\nu)},N_\nu)
\frac{dE_{k'}}{D_i},
\end{eqnarray}
where $\Delta\equiv E_\nu-E_i-\omega_{ab,ch} $, is a ``spread'' delta-function
which was studied in Refs. \cite{Flambaum:93,FG95,Ce}. It peaks at $\Delta =0$
and describes the approximate energy conservation for the transition between
compound states induced by the two-electron transition $ch\rightarrow ab$
and broadened by the spreading widths. For the Breit-Wigner strength functions
one has
\begin{equation}\label{spreading-delta}
\tilde{\delta}(\Gamma_{\rm spr}^{(i)},\Gamma_{\rm spr}^{(\nu)},\Delta)=
\frac{1}{2\pi}\frac{\Gamma_{\rm spr}}{\Delta^2+\Gamma_{\rm spr}^2/4},
\end{equation}
where $\Gamma_{\rm spr}=\Gamma_{\rm spr}^{(i)}+\Gamma_{\rm spr}^{(\nu)}$.
From Eqs. (\ref{eq:MSM}) and (\ref{mean-weight4}), the mean-squared matrix
element of the two-body operator between the compound states is finally obtained
as
\begin{eqnarray}\label{eq:MSM-final}
\overline{|\langle \Psi _\nu |\hat V |\Psi_i\rangle |^2} =
\frac{1}{4}\sum_{(ab)ch} | \langle ab |\hat v |hc\rangle
-\langle ba |\hat v |hc\rangle |^2 \nonumber \\
 \times  \langle\hat{n}_h\hat{n}_c(1-\hat{n}_a)(1-\hat{n}_b)\rangle_i D_\nu
\tilde{\delta}(\Gamma_{\rm spr}^{(i)},\Gamma_{\rm spr}^{(\nu)},\Delta).
\end{eqnarray}
In this expression the summation is carried out over the single-electron
states $a$, $b$, $h$, and $c$. Note that if $|\Psi_i\rangle $ is a simple,
unmixed state, there is no sum over $k'$ in
Eqs.~(\ref{mean-weight})--(\ref{eq:emptiness}). In this case
$\Gamma_{\rm spr}^{(i)}=0$ and $\Gamma_{\rm spr}=\Gamma_{\rm spr}^{(\nu)}$
in Eq.~(\ref{spreading-delta}).

For a one-body operator $\hat M=\sum_{ab}M_{ab}a^{\dagger}_{a}a_b$, the
mean-squared matrix element is obtained similarly
\cite{Flambaum:93,FG95,Ce}:
\begin{eqnarray}\label{eq:MSM-final2}
\overline{|\langle \Psi _\nu |\hat M |\Psi_f\rangle |^2} =\sum_{ab}
|\langle a|\hat m|b\rangle|^2\langle\hat{n}_a(1-\hat{n}_b)\rangle_\nu
\nonumber \\ 
\times D_f \tilde{\delta}(\Gamma_{\rm spr}^{(\nu)},\Gamma_{\rm spr}^{(f)},
E_f-E_\nu-\omega_{ba}),
\end{eqnarray}
where $\omega_{ba}=\varepsilon_b-\varepsilon_a$ is the energy of the
single-electron transition $a\rightarrow b$.

\subsection{Capture cross section and autoionization width}
\label{subsec:aw}
The autoionization width
$\Gamma_\nu^{(a)}=2\pi|\langle \Psi _\nu |\hat V |\Psi_i \rangle |^2 $ gives the
transition rate between the initial state, $e^-+A^{q+}$, and the
multiply exited compound resonance of the ion $A^{(q-1)+}$ due to the two-body
Coulomb interaction $\hat V $. Unlike the complex multiply excited states
$|\Psi _\nu\rangle $, the initial state of the recombination process is simple.
It describes an electron with the energy $\varepsilon $ incident on the
ground (or low-lying excited) state $|\Phi _0\rangle$ of the target, which is
often dominated by one configuration. It is clear that the autoionization width
averaged over compound resonances is determined by the mean-squared matrix
element of the Coulomb interaction between electrons, which is given by
Eq.~(\ref{eq:MSM-final}). The initial state 
$|\Psi _i\rangle=|\Phi _0,c\rangle$ is thus a compound state with
negligible spreading width $\Gamma_{\rm spr}^{(i)} \ll \Gamma_{\rm spr}^{(\nu)}$.
The total width of the function $\tilde \delta $, Eq.~(\ref{spreading-delta}),
is dominated by the compound resonance width
$\Gamma_{\rm spr}\approx \Gamma_{\rm spr}^{(\nu)}$.

Nonzero contributions to the sum in Eq. (\ref{eq:MSM-final}), i.e.,
to $\Gamma^{(a)}$, arise from the basis states which differ from the initial
state $|\Phi _0,c\rangle $ by the single-particle states of two
electrons. Therefore, it is sufficient to sum over the doubly-excited basis
states,
$$\Gamma^{(a)}=2\pi\sum_d \overline{C_d^{(\nu)2}} |\langle \Phi _d |\hat V |\Phi _0,c \rangle |^2. $$ 
Such two-electron excitations $|\Phi _d\rangle$ play the role of
{\em doorway states} for the electron capture process. Since these states are
not the eigenstates of the system they have a finite energy width
$\Gamma _{\rm spr}$. The wave function of a doorway state can be constructed
using the creation-annihilation operators, $|\Phi _d\rangle =a_a ^\dagger
a_b ^\dagger a_h |\Phi _0,c\rangle $, where $a \equiv n_a l_a j_a  m_a $ and
$b \equiv n_b l_b j_b m_b $ are excited single-electron states, and
$h \equiv n_h l_h j_h m_h $  corresponds to the hole in the target ground state.
Of course, to form the doorway states with a given total angular momentum
$J$, the excited electrons and the ionic residue must be coupled into $J$.
However, the $2J+1$ factor and summation over $J$ in Eq.~(\ref{aCapcs}) account
for all possible couplings. This means that the sum over the eigenstates in
Eq.~(\ref{aCapcs}) can be replaced by the sum over the one-hole-two-electron
excitation, as in Eq. (\ref{eq:MSM-final}), and one obtains the capture cross
section in the form of Eq.~(\ref{sigma}).

Note that when the number of active electrons and orbitals is large,
the occupation numbers for different orbitals become statistically independent.
In this case, the correlated product of the single-particle occupancies,
Eq.~(\ref{eq:emptiness}) can be approximated by the fractional occupation
numbers of the electronic {\em subshells} with definite $j$.
The orbital $c$ is
taken a continuum, $c \equiv \varepsilon ljm$ in Eq. (\ref{sigma}). Its
wave function is normalized to the delta-function of energy, and it is
occupied in the initial state, i.e., $\hat n_c=1$. After summation over the
magnetic quantum numbers $m_a$, $m_b$, etc., and angular reduction of the
Coulomb matrix elements, the final expression for the capture cross section is
\begin{eqnarray}\label{eq:capture-explicit}
\bar \sigma _c & = &\frac{\pi^2}{k^2}\sum _{abh ,lj}
\frac{\Gamma _{\rm spr}}{(\varepsilon -\varepsilon _a -
\varepsilon _b +\varepsilon _h )^2 +\Gamma _{\rm spr}^2/4}
\nonumber\\
 &\times &\sum _\lambda \frac{\langle a ,b \| V_\lambda \| h ,
\varepsilon lj \rangle }{2\lambda +1}
\Biggl[ \langle a ,b \| \hat V _\lambda \| h ,
\varepsilon lj\rangle  - (2\lambda +1) \nonumber \\
&\times &\sum _{\lambda '}
(-1)^{\lambda +\lambda '+1}\left\{ {\lambda \atop \lambda '}{j_a 
\atop j_b }{j \atop j_h }\right\}
\langle b ,a \| \hat V_{\lambda '}\| h,\varepsilon lj\rangle \Biggr]\nonumber \\
&\times & \frac{n_h }{2j_h +1}\left(1-\frac{n_a }{2j_a 
+1}\right)\left(1-\frac{n_b}{2j_b  +1}\right).
\end{eqnarray}
Here $n_a$, $n_b$ and $n_h$ are the occupation numbers of the corresponding
subshells (ranging from 0 to $2j_a+1$, etc.), and $\varepsilon _a $,
$\varepsilon _b $, and $\varepsilon _h $ are their energies. The two terms in
square brackets represent the direct and exchange contributions, and
$\langle a ,b \| V_\lambda \| h ,\varepsilon lj \rangle $ is the reduced
Coulomb matrix element:
\begin{eqnarray}\label{eq:redmael}
\langle a ,b \| V _\lambda \| h ,c \rangle &=& 
\sqrt{(2j_a+1)(2j_b+1)(2j_h+1)(2j_c+1)}\nonumber \\
&\times &\xi(l_a +l_c +\lambda )\xi(l_b +l_h +\lambda )\\
&\times & \left(  {\lambda \atop 0}{j_a \atop -\frac{1}{2}}
{j_c \atop \frac{1}{2}}\right) \left( {\lambda \atop 0}
{j_b \atop -\frac{1}{2} }{j_h \atop \frac{1}{2}}\right) R_\lambda (a ,b ;h ,c ),
\nonumber
\end{eqnarray}
where $\xi (L)=[1+(-1)^L]/2$ is the parity factor, and
\begin{eqnarray}\label{eq:Radint}
R_\lambda (a ,b ;h ,c ) 
 &=&  \iint \frac{r_<^\lambda }
{r_>^{\lambda +1}}[f_a (r)f_c (r)+g_a (r)g_c (r)] \nonumber\\
&\times &[f_b (r')f_h (r')+g_b (r')g_h (r')]drdr'
\end{eqnarray}
is the radial Coulomb integral, $f$ and $g$ being the upper and lower 
components of the relativistic orbital spinors.

Once $\bar \sigma _c$ is known, Eq.~(\ref{aCapcs}) allows one to estimate the
average ratio $\Gamma^{(a)}/D$ for a typical $J^\pi$,
\begin{equation}\label{aCapcs2}
\left\langle \frac{\Gamma^{(a)}}{D}\right\rangle =\frac{k^2(2J_i+1)\bar\sigma_c}
{\pi^2\sum_{ J^\pi}(2J+1)}=\frac{k^2(2J_i+1)\bar \sigma _c}{2\pi^2 J_{\rm max}^2}
\end{equation}
where the sum in the denominator is over the angular momentum and parity
$J^\pi$ which contribute effectively to the capture cross section. For example,
$J_{\rm max}\approx 10$ and $J_i=6$ for the recombination of Au$^{25+}$ and
W$^{20+}$. A typical distribution of level densities $\rho_{J^\pi}$ for
different $J$, is shown on the inset of Fig.~\ref{fig:density}.

\subsection{Radiative width}\label{subsec:rad}

The second step of the recombination process is radiative stabilization. Any
excited electron in the compound state $|\Psi _\nu \rangle $ can emit a photon.
Using Eq.~(\ref{eq:MSM-final2}) the total photoemission rate $\Gamma ^{(r)}$
can be estimated as a weighted sum of the single-particle rates,
\begin{equation}\label{eq:Gamma_r}
\Gamma ^{(r)}\simeq \sum _{a ,b }
\frac{4\omega _{b a }^3} {3c^3}
|\langle a \|\hat d\|b \rangle |^2
\left\langle \frac{n_b }{2j_b +1}\left( 1-\frac{n_a }
{2j_a +1}\right) \right\rangle _\nu ,
\end{equation}
where $\omega _{b a }=\varepsilon _b -\varepsilon _a >0$,
$\langle a \|\hat d\|b \rangle $ is the reduced dipole operator
between the orbitals $a $ and $b $, and $\langle \dots \rangle _\nu $ is the
mean occupation number factor. The mean subshell occupation numbers for a
given energy can be obtained by averaging over the basis states involved, e.g.,
\begin{equation}
n_a(E)=\sum_k \overline{C^2_k}(E)n^{(k)}_a,
\end{equation}
where $n_a^{(k)}$ is the occupation number of the subshell $a$ in the basis
state $k$.

Since $|\Psi _\nu \rangle$ have large numbers of principal components $N$,
the fluctuations of their radiative widths are small, $\sim 1/\sqrt{N}$.
This can also be seen if one recalls that a chaotic
multiply excited state is coupled by photoemission to many lower-lying states,
and the total radiative width is the sum of a large number of (strongly
fluctuating) partial widths. A similar effect is known in compound nucleus
resonances in low-energy neutron scattering \cite{Bohr:69, FS84}.
In multicharged ions with dense spectra of chaotic multiply-excited states,
the autoionization widths are suppressed as $\Gamma^{(a)}\propto 1/N $.
Physically this happens because the coupling strength of the two-electron
doorways state to the continuum is shared between many complex
multiply-excited eigenstates. The radiative width does not have this
suppression since all components of a compound state contribute to the
radiative decay. As a result, the radiative width may dominate in the total
width of the resonances, $\Gamma ^{(r)} \gg \Gamma ^{(a)}$, making their
fluorescence yield close to unity. Our numerical results for the
recombination of Au$^{25+}$ presented in \cite{Flambaum02}, supported
this picture.

\section{Numerical results}\label{sec:num}

In this section we apply our theory to calculate the recombination rate for
the tungsten ions from W$^{17+}$ to W$^{24+}$. Experimental data are
available for the recombination of W$^{20+}$ forming
W$^{19+}$~\cite{schippers11}. We will use this system as an example to
describe the calculations. Calculations for other ions are similar.

\begin{figure}[t!]
\centering
\includegraphics*[width=0.48\textwidth]{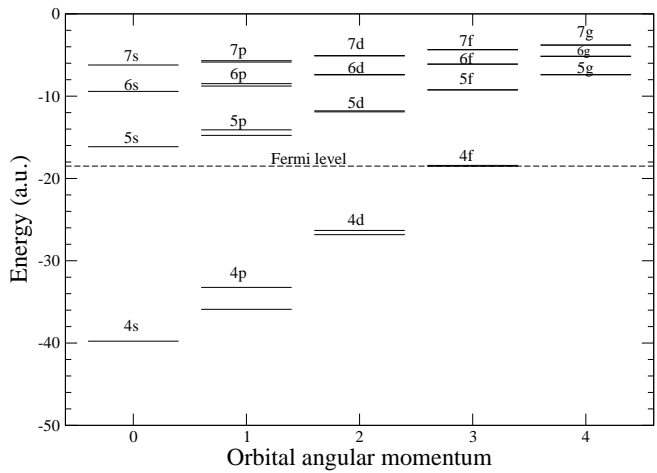}
\caption{Energies of the occupied and vacant single-particle
orbitals of W$^{19+}$ obtained in the Dirac-Fock calculation.}
\label{fig:orbitals}
\end{figure}

When an electron recombines with W$^{20+}$, it can be captured into an excited
state of the compound W$^{19+}$ ion. Its ground state belongs to the
$1s^2\dots 4f^9$ configuration.
Figure \ref{fig:orbitals} shows the energies of its relativistic orbitals
$nlj$ obtained in the Dirac-Fock calculation. All orbitals below the Fermi
level, $1s$ to $4f$, were obtained in the self-consistent calculation of the
W$^{19+}$ ground state. Each of the excited-state orbitals above the Fermi
level: $5s$, $5p$, etc., was calculated by placing one electron into it, in
the field of the frozen W$^{20+}~1s^2\dots 4f^8$ core. The energy of the
highest orbital occupied (partially) in the ground state is
$\varepsilon _{4f_{7/2}}=-18.41$ a.u. This value gives an estimate of the
ionization potential of W$^{19+}$: $I\approx |\varepsilon _{4f_{7/2}}|=18.41$~a.u.
This value is in agreement with NIST data, $I = 18.47$~a.u.\cite{NIST}.

Excited states of the ion are generated by transferring one, two,
three, etc. electrons from the ground-state orbitals into the empty orbitals
above the Fermi level (Fig. \ref{fig:orbitals}), or into the partially
occupied $4f$ orbitals. We are interested in the excitation spectrum of
W$^{19+}$ near its ionization threshold. This energy ($\sim $20 a.u.)
is sufficient to push up a few of the nine $4f$ electrons, and even excite
one or two electrons from the $4d$ orbital. However, the
preceding $4p$ orbital is already deep enough to be considered inactive.
Thus, we treat W$^{19+}$ as a system of 19 electrons above the frozen
Kr-like $1s^2\dots 4p^6$ core. Note that in constructing the excited state
configurations, we disregard infinite Rydberg series which 
correspond to the excitation of one electron in the field of W$^{20+}$.
Rydberg states belong to a single-particle aspect of the
$e^-+\mbox{W}^{20+}$ problem, and are not expected to contribute
much to the recombination cross section in this system.

Assuming that the fluorescence yield is close to unity (see below), we
calculate the recombination cross section from Eq.~(\ref{eq:capture-explicit}).
Before using this formula, one needs to obtain a list of two-electron-one-hole
excitations of W$^{19+}$ with energies close to the ionization threshold,
which act as the doorway states. One also needs to estimate the
spreading width $\Gamma _{\rm spr}$. For low-energy electron recombination,
we restrict the consideration to the energy interval,
\begin{equation}
E = I \pm \Delta E/2,
\label{eq:E}
\end{equation}
where $E=0$ for the ground state of the final-state ion, and
we choose $\Delta E\sim \Gamma _{\rm spr}$. In practice, we start from some
initial estimate of the spreading width, and subsequently find a more
accurate value using an iterative procedure.

The spreading width is found from Eq.~(\ref{gammaspread}) by constructing the
Hamiltonian matrix for a number of configurations within the interval
(\ref{eq:E}), and averaging the squared offdiagonal matrix elements of the
Hamiltonian matrix $H_{ij}$,
\begin{equation}
\overline{H^2_{ij}}  = \frac{2}{N_s(N_s-1)}\sum_{i<j} H^2_{ij},
\label{eq:H2}
\end{equation}
where the average is taken over $N_s$ basis states whose energies
$E_k\equiv H_{kk}$ lie within the energy interval (\ref{eq:E}). The level
spacing $D$ in Eq.~(\ref{gammaspread}) is found as average energy interval
between the states
\begin{equation}
D = \Delta E/N_s.
\label{eq:D}
\end{equation}
The list of two-electron-one-hole excitations $h^{-1}ab$ which contribute to
the sum (\ref{eq:capture-explicit}) is found by checking which of these
configurations contribute basis states to the sum in Eq.~(\ref{eq:H2}).
It is known that the spreading width is a robust characteristic of the
system. Indeed, we have checked that when more configurations are included,
both $D$ and $\overline{H^2_{ij}}$ decrease, whereas $\Gamma_{\rm spr}$ does
not change much (see also Ref.~\cite{Gribakin03}).

When finding $\overline{H^2_{ij}}$ and $D$ we use basis states with definite
projection of the total angular momentum $J_z$ corresponding to the minimal
value of $J_z$ (0 or $1/2$), rather than the states with definite total
angular momentum $J$. This method is significantly simpler than the use of
the basis states with definite $J$ and $J_z$, and produces the same results
for $\Gamma_{\rm spr}$, Eq.~(\ref{gammaspread}).

Table \ref{t:W} shows the spreading widths for the compound ions of
tungsten, W$^{(q-1)+}$, with excitation energies close to the ionization
threshold, formed in the process of low-energy electron recombination with
W$^{(q)+}$.
With the exception of the target ion with the smallest number of $4f$ electrons
(W$^{25+}4f^3$), the spreading widths are in the range 0.5--0.7~a.u. In fact,
the value of $\Gamma _{\rm spr}$ does not strongly affect the magnitude
of the capture cross section, Eq.~(\ref{eq:capture-explicit}), since the
area under the Breit-Wigner contour corresponding to each doorway $h^{-1}ab$
is independent of $\Gamma _{\rm spr}$.

\begin{table}[t]
\caption{Electron capture cross sections $\bar \sigma _c $ and rate
coefficients $\alpha _c$ for the tungsten ions W$^{(q)+}$ with the open
$4f$ subshell, and properties of the compound ions  W$^{(q-1)+}$ at excitation
energies close to the ionization threshold $I$.}
\label{t:W}
\begin{ruledtabular}
\begin{tabular}{lcccccc}
Target & $I$\footnotemark[1] & $D$ & $K$ & $\Gamma_{\rm spr}$ &
$\bar \sigma _c$\footnotemark[2]&$\alpha_c$\footnotemark[2] \\
~~ion    & a.u. & $10^{-4}$\,a.u.& & a.u. & $10^{-16}\,{\rm cm}^2$&
$10^{-7}\,{\rm cm}^3/\mbox{s}$ \\
\hline \\[-9pt]
W$^{18+}4f^{10}$ & 15.5 & 0.2 & 70 & 0.56  & 25 & 1.5  \\
W$^{19+}4f^9$ & 17.0 & 0.1 &  93 & 0.65  & 29 & 1.7  \\
W$^{20+}4f^8$    & 18.5 & 0.1 & 105 & 0.68  & 30 & 1.8  \\
W$^{21+}4f^7$    & 20.0 & 0.1 &  96 & 0.68  & 34 & 2.0  \\
W$^{22+}4f^6$    & 21.8 & 0.2 &  76 & 0.65  & 16 & 0.98  \\
W$^{23+}4f^5$    & 23.5 & 0.4 &  48 & 0.59  & 11 & 0.67  \\
W$^{24+}4f^4$    & 25.2 & 1.3 &  25 & 0.50  & 19 & 1.1  \\
W$^{25+}4f^3$    & 27.0 &  11 &   5 & 0.16  & 12 & 0.7  \\ 
\end{tabular}
\footnotetext[1]{Ionization energy of the final-state ions, Ref.~\cite{NIST}.}
\footnotetext[2]{Capture cross section from Eq.~(\ref{eq:capture-explicit})
and rate coefficient for incident electron energy $\varepsilon =1$~eV.}
\end{ruledtabular}
\end{table}

As discussed in Sec.~\ref{subsec:chaos}, strong mixing of the basis states 
results in the eigenstates with large numbers of principal components,
$N\sim \Gamma _{\rm spr}/D\sim \overline{H^2_{ij}}/D^2\gg 1$. This
occurs when
\begin{equation}
K=\sqrt{\overline{H^2_{ij}}}/D \gg 1.
\label{eq:C}
\end{equation}
Table \ref{t:W} shows that this criterion is fulfilled for all the ions
studied, and that the expected number of principal components is indeed large,
$N\sim 10^4$. As explained in Sec.~\ref{subsec:rad}, in this case one
can expect large fluorescence yields, $\omega_f\approx 1$. This means that the
recombination cross section will be at the limit given by the total electron
capture cross section, Eq.~(\ref{eq:capture-explicit}).

In the present calculations of $\bar \sigma _c$,
Eq.~(\ref{eq:capture-explicit}), we also include in a semiempirical way the
effect of screening of the Coulomb interaction between valence
electrons by core electrons. This is done by introducing the screening
factors $f_\lambda $ in the two-electron Coulomb integrals, assuming that these
factors depend on the Coulomb integral multipolarity $\lambda $ only. The
factors were calculated to be $f_1=0.7$, $f_2=0.8$, $f_3=0.9$ \cite{Dzuba08}.
Coulomb integrals of other multipolarities are not modified. The above values
of the screening factors were found in the calculations for other atomic
systems. However, in practice they change little from one atom to another.

\begin{figure}
\centering
\includegraphics*[width=0.48\textwidth]{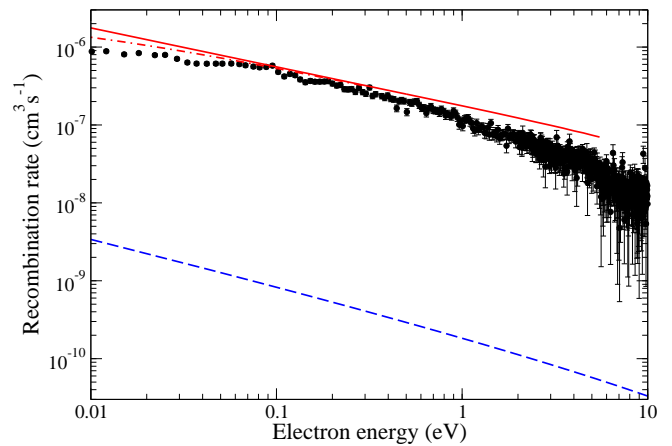}
\caption{Recombination rate coefficient of W$^{20+}$. Dashes line is the direct
radiative recombination rate, Eq. (\ref{eq:sigmad}); solid line is the
capture rate calculated using the present theory,
Eq.~(\ref{eq:capture-explicit}); dot-dashed line shows the same, taking into
account the velocity distribution of the electron beam (see
text); solid circles is the measured rate coefficient~\cite{schippers11}.}
\label{fig:rate}
\end{figure}

To compare with experiment for W$^{20+}$ \cite{schippers11}, the cross section
obtained from Eq.~(\ref{eq:capture-explicit}) is converted into the
rate coefficient $\alpha _c=\bar \sigma _c v$, where $v$ is the velocity of
the incident electron. The result is shown in Fig.~\ref{fig:rate} by the solid
line. Since the sum in Eq.~(\ref{eq:capture-explicit}) has a weak dependence
on the electron energy, the capture cross section at low energies is
proportional to $1/\varepsilon $, and the corresponding rate coefficient
behaves as $\alpha _c\propto 1/v$. The calculated rate agrees well with the
experimental data in the energy range of 0.1--1 eV. At higher energies the
experimental rate coefficient tends to drop faster than $1/v$.

Figure \ref{fig:rate} also shows the rate coefficient for the direct
radiative recombination. The latter is estimated using the Kramers formula
for the radiative recombination cross section \cite{Sobelman92} (in
atomic units),
\begin{equation}\label{eq:sigmad}
\sigma_d^r= \frac{32\pi }{3\sqrt{3}c^3}\,\frac{Z_i^2}
{k^2} \ln \left( \frac{Z_i}{n_0k}\right) ,
\end{equation}
where $Z_i$ is the ionic charge $Z_i$ (e.g., $Z_i=20$ for $e^-+W^{20+}$),
and $n_0$ is the principal quantum number of the lowest unoccupied ionic
orbital ($n_0=5$ for W$^{20+}$) \cite{Gribakin99}. The energy dependence of
this cross section is close to $1/\varepsilon $, and the corresponding
rate coefficient (dashed line in Fig.~\ref{fig:rate}) is three orders of
magnitude smaller than the measurement in the energy range shown.

Below $\varepsilon =0.1$~eV the measured recombination rate coefficient can
be affected by the velocity distribution of the electron beam, which is
characterized by two temperatures, $T_\parallel = 0.15$~meV and
$T_\perp = 10$~meV \cite{schippers11}. Taking this into account (see Eq.~(18)
in Ref.~\cite{Flambaum02}) reduces the calculated resonant capture rate below
50 meV (dot-dashed line in Fig.~\ref{fig:rate}), bringing it into closer
agreement with experiment.

As discussed above, the capture cross section has a simple $1/\varepsilon $
energy dependence at low electron energies. Hence, in Table~\ref{t:W}
we show the cross sections and rate coefficients for W$^{q+}$ ($q=18$--25)
calculated at one low electron energy, $\varepsilon =1$~eV.
We see that the largest cross section is predicted for the ion with the
half-filled $4f$ subshell. On the other hand, all the cross sections
are within a factor of three of each other, and much larger than what
one would expect from the direct RR process, Eq.~(\ref{eq:sigmad}).

Of course, one must keep in mind that compared with the capture cross section,
the recombination cross section contains an additional factor $\omega_f$. 
The fluorescence yield may be significantly smaller than $\omega_f=1$ for ions,
in which the degree of mixing is not as large as it is in the compound W$^{19+}$
ion. In particular, this may be the case for the ions in which the mixing
strength $K$ (see Table \ref{t:W}) and the number of principle components
$N$ is not too large. In this case one should regard $\bar \sigma _c$
as the upper limit, and use Eqs.~(\ref{radiative-final}) and (\ref{eq:rec_app})
to estimate the recombination cross section.

\section{Conclusions}

A detailed derivation of the statistical theory of resonant electron capture 
by many-electron ions has been presented. Numerical calculations have been
performed for a number of tungsten ions with a partially filled $4f$
subshell. The calculated rate coefficient for W$^{20+}$ is in agreement with the
measurements at low electron energy. The present approach can be used to
investigate other processes mediated by chaotic, multielectronic excited
states.

After completing the present work, we became aware of Ref.~\cite{Badnell12}
which considers dielectronic recombination of W$^{20+}$.

\section*{Acknowledgments}
This work was funded in part by the Australian Research Council. We
thank S. Schippers for providing the experimental data in numerical form,
and acknowledge helpful conversations with J. Berengut. GG is grateful to the
Gordon Godfrey fund (UNSW) for support.


\end{document}